# Optimization of Financial Instrument Parcels in Stochastic Wavelet Model

A.M.Avdeenko

aleksei-avdeenko@mail.ru

## 1. Introduction

Exchange rate fluctuation, values of shares, futures and options are random or quasirandom processes, and many "objective" (or perceiving as such) and random or seeming (perceiving) as random factors influence on them.

Here a number of questions arise. Whether it is possible on the basis of analysis of limited numerical sequence of financial instrument cost to draw a conclusion about the nature of statistical processes underlying the observed event, specifically, to decide a question about stationarity (quasi-stationarity) of the process, to restore probability density upon observed data, to calculate probability of certain events etc.?

Preceding the result of suggested work, it is necessary to point out that the answer is positive, specifically, within the framework of not so critical axiomatics, creation of desired statistics on the basis of analysis of random or quasirandom sequences of financial instrument costs is possible.

Furthermore, according to the author, solution of the problem is essentially eased because we do not speak about simulation of complicated natural processes, but only about simulation of decision making method used by market participants, who have the same initial information and comparable computational and analytical capabilities, which pursue well-defined and, generally speaking, obvious target.

### 2. Model of nonlinear stochastic wavelets (NSW model)

Let random process $X(t,\omega)$ be the cost of shares, options and futures in the point of time $t \subset R$, where $\omega \subset \Omega$ is a point in probability space with measure function $P$. Let's introduce a population of analyzing functions-wavelets $\psi_{i\tau}(t) = 2^{-i/2}\psi(2^{-i}t - \tau)$, meeting the condition $\psi(t) \subset L^1 \cap L^2$, $\int \psi(t)dt = 0$ and forming the orthonormal basis for $L^2(R)$ with compact carrier. Small value of index $j$ conforms to low-frequency component of the random function $X(t,\omega)$, and the large one to high-frequency component [1].

Let's express the random function in the form of linear combination of functions $Y_i(t) = \int X(t,\omega)\psi_{i\tau}(t)dt$. We designate $\mathbf{Y}(t) = (Y_1(t),..Y_J(t))$, where - $J$ is maximum order of wavelet transformation. We presume that new $J$ - dimensional random vector $\mathbf{Y}(t)$ is in accord with stochastic equation of Ito:

$$d\mathbf{Y}(\tau) = \mathbf{F}(\mathbf{Y}(\tau))d\tau + \mathbf{G}(\mathbf{Y}(\tau)) \cdot d\boldsymbol{\omega}, \tag{1}$$

where $d\omega_i(t)$ is random Wiener process $< d\omega_i(t) >= 0, \quad < d\omega_i(t) d\omega_j(t_1) >= \delta_{ij}\delta(t - t_1),\ i, j = 1...J$.

Symbol $<...>$ means averaging as per random process $X(t,\omega)$ with measure function $P$, and $\mathbf{F}(\mathbf{Y}(\tau)), \mathbf{G}(\mathbf{Y}(\tau))$ are drift and diffused components accordingly, we assume that $G_{ij}(\mathbf{Y}(\tau)) = \delta_{ij} G(\mathbf{Y}(\tau))$.

Conventional transition probability $f(\mathbf{y}(t)/\mathbf{y}(t_0))$ can be represented as a path integral [2]. Nonlinear effects can be taken into account using diagrammatic method. Specifically, we can say that for nonlinear model the spectral potency (Fourier transform of two-point twin correlation function), some times has power asymptotics, i.e. observed "fractalness" of signal is only the result of nonlinearity of the model.

In single-mode case $J = 1$, time-independent density of distribution defined as $P(Y_1 < y_1) = \int_{-\infty}^{y_1} f_s(y_1) dy_1$ can be represented in the form of quadratures: $f_s(y_1) \approx e^{W(y_1)}$, where $W(y_1) = \int_0^{y_1} \frac{2F_1(y_1)}{G_1^{\ 2}(y_1)} dy_1$.

In general $J$ - dimensional case fulfillment of additional requirement, i.e. condition of detailed balance [3], is required and it is natural. However, if stationary solution exists, then it always can be represented in exponential form $f_s(\mathbf{y}) \approx e^{W(\psi_k(\mathbf{y}))}$, where $\psi_1(\mathbf{y})...\psi_k(\mathbf{y})$ are certain functions of random variables $\mathbf{y}$. This result can be received based on the principle of maximum informational entropy [3] or using inequation of Rao-Cramer. In this case the moments $< \psi_1(\mathbf{y})...\psi_k(\mathbf{y}) >$ are limitations (links), under which we find extremum $< \ln f_s(\mathbf{y}) >$.

Let $H_k(x_n)$ be adequate set of orthogonal functions, for example, Ermiht polynoms of $k$ order. Orthogonal decomposition $\mathbf{F}(\mathbf{Y})$, $\mathbf{G}(\mathbf{Y})$ per polynoms $H_k(Y_n)$ is given by [2]:

$$\mathbf{F}(\mathbf{Y}) = \sum_{n...p} \sum_{k...l} \lambda_{k...l}^{n...p} H_k(Y_n)...H_l(Y_p)$$
$$G_{jj}(\mathbf{Y}) = \sum_{n...p} \sum_{k...l} Q_{k...l,j}^{n..p} H_k(y_n)...H_l(y_p) \tag{2}$$

Aim of further stochastic simulation is reconstruction of motion equation (1) and conditional probabilities within the framework of made suppositions based upon available observations (experimental information) or, that is equivalent, determination of decomposition values in proportion (2).

To solve the problem we shall use extremal principle, i.e. we shall define the value $\lambda_{k...l}^{n...n-p}$, $Q_{k...l,j}^{n...n-p}$ in such a manner that minimize functionals that are norms of differences of observed and simulated values averaged per random process with measure function $P$. After differentiating these expressions with respect to $\lambda_{k...l}^{n...n-p}$, $Q_{k...l,j}^{n...n-p}$, multiplying by $H_k(y_n)$ and integrating on $y_n ... y_p$ we have two systems of linear equations for unknown variables $\lambda_{k...l}^{n...n-p}$, $Q_{k...l,j}^{n...n-p}$.

Solution of these systems gives us expression for drift and diffused factors of non-Markov stochastic equation, i.e. the most probable path function of the process and statistic of fluctuations round it. We shall name corresponding sequence of random variables as setting up sequence of $T_0$ length.

On the basis of these results we can construct optimal policy of portfolio management. It follows from compactedness and condition $\int \psi(t)dt = 0$ that factors of wavelet decomposition characterize mean change of value $X(\tau,\omega)$ on the length of carrier $2^{-j}\tau_0$.

Therefore, in elementary case, moment of optimal purchase of securities can be represented as $t = \inf(-dy_1 > 0, P_s > 1-\alpha_1)$ and moment of optimal sale as $t = \inf(-dy_1 < 0, P_s < \alpha_1)$, where $P_s = \int_{-\infty}^{0} f_s(y)dy$. Sign of $dy_1$ has been chosen for Haar wavelet, and value is risk level. The model under consideration has non-Markov nature.

Conditions $P_s > 1-\alpha_1, P_s < \alpha_1$ are required for exception (on the average) of the purchase in case of $Y_1(\tau_{n-1}) < Y_1(\tau_n), -dY_n < 0$ and sale in case of $Y_1(\tau_{n-1}) > Y_1(\tau_n), -dY_n > 0$ .

Alternative variant is use of convolution $f_s(z) = \int_{-\infty}^{+\infty} f_s(y_1) f_{s,T}(y_1+z)dy_1$. In this case criteria $P_s > 1-\alpha_1, P_s < \alpha_1$ mean resale- or repurchase moments.

In both cases it is necessary to exclude statistically nonsignificant differences between stationary distributions in time displacement $t \to t+T$ .

The last is possible when using non-parametric Kolmogorov criterion with risk level $\alpha_2$, i.e. synthesized stationary distributions $f_s'(y_1), f_s'(y_1)$ in displacement $t \to t+T$ with probability $p = 1-\alpha_2$ are considered to be indistinguishable, if $\max\left|K(y_1(t_n)-K'(y_1(t_n))\right| < k(\alpha_2)/\sqrt{N}$, where $K, K'$ are cumulative probabilities for this condition and condition received as a result of displacement $T$ .

The constant $k(\alpha_2) \approx 1$ depends on risk level $\alpha_2$ , $N$ is the number of experimental points (samples) per which approximation of stationary distribution has been constructed. Conditions of purchase/ sale of financial instruments in NSW model are shown schematically in figure 1.

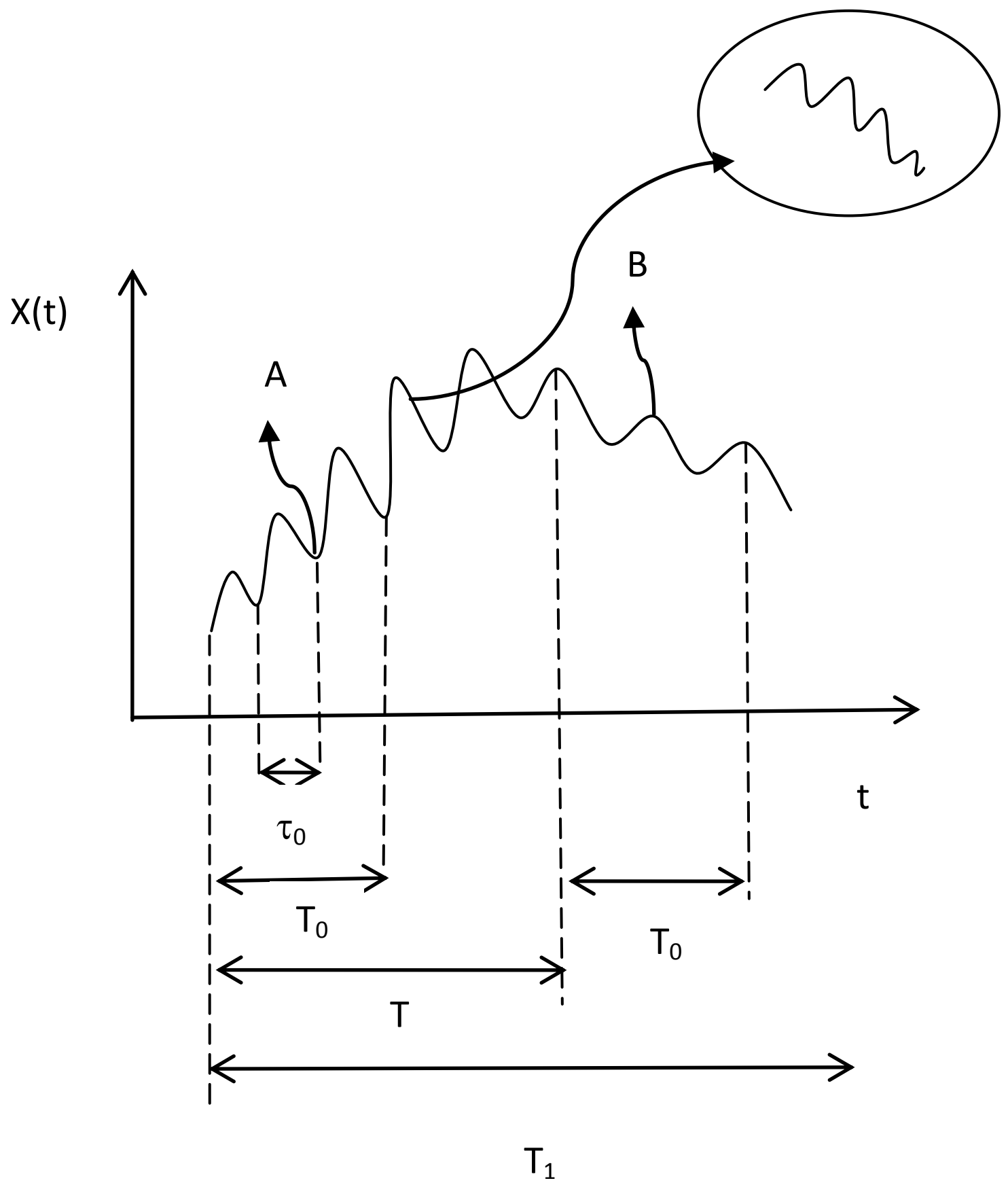

Figure 1. Conditions of purchase/ sale of financial instruments in NSW model; here $\tau_0$ is compactedness interval of analyzing wavelet, $T_0$ is length of setting up sequence, $T$ is interval of quasi-stationarity of distribution with risk level $\alpha_1$, $T_1$ is interval of parcel redistribution decision making, and $A, B$ are points of purchase and sale of financial instruments.

### 3. Experimental proof of NSW model

Data of three types that are stock prices of Russian emitters (Sberbank, SurgutNefteGas and Gasprom) and stocks of U.S. companies (AT&T Inc, Dell Inc. and the Bank of America) over a period since January 1, 2009 till December 31, 2009 were used as initial database for model run.

It was analyzed the non-Markov dual-mode model $(J = 2)$, scanning was implemented in increments of 1 min., calibration sequence amounts to 32-64 indications, and the order of Hermite polynom not to exceed $K = 3$.

Analyzing function - Haar wavelets $\phi_H$, Daubechies wavelets $\phi_2, \phi_3$ and Bettle-Lamerie spline wavelets $B_1, B_2, B_3$ [1]. Although the last ones do not have compact carrier, they come down quickly enough under large time.

Cost of securities at the start time was assumed as $C(t_0) = 1$. Objective function of the model is profit at the point of time $t$: $Z(t) = C(t) / C(t_0)$, decision making for purchase/ sale in accordance with proposed criterion and calculation algorithm for values $\mathbf{F}(\mathbf{Y}), \mathbf{G}(\mathbf{Y})$ at the point of time $t > t_0$ was implemented in automatic mode based upon prior information. Risk levels were taken as $\alpha_1 = \alpha_2 = 0.05$.

Results of calculation for three groups of active securities are given in figure 1. The results are designated at intervals of five acts of purchase and sale.

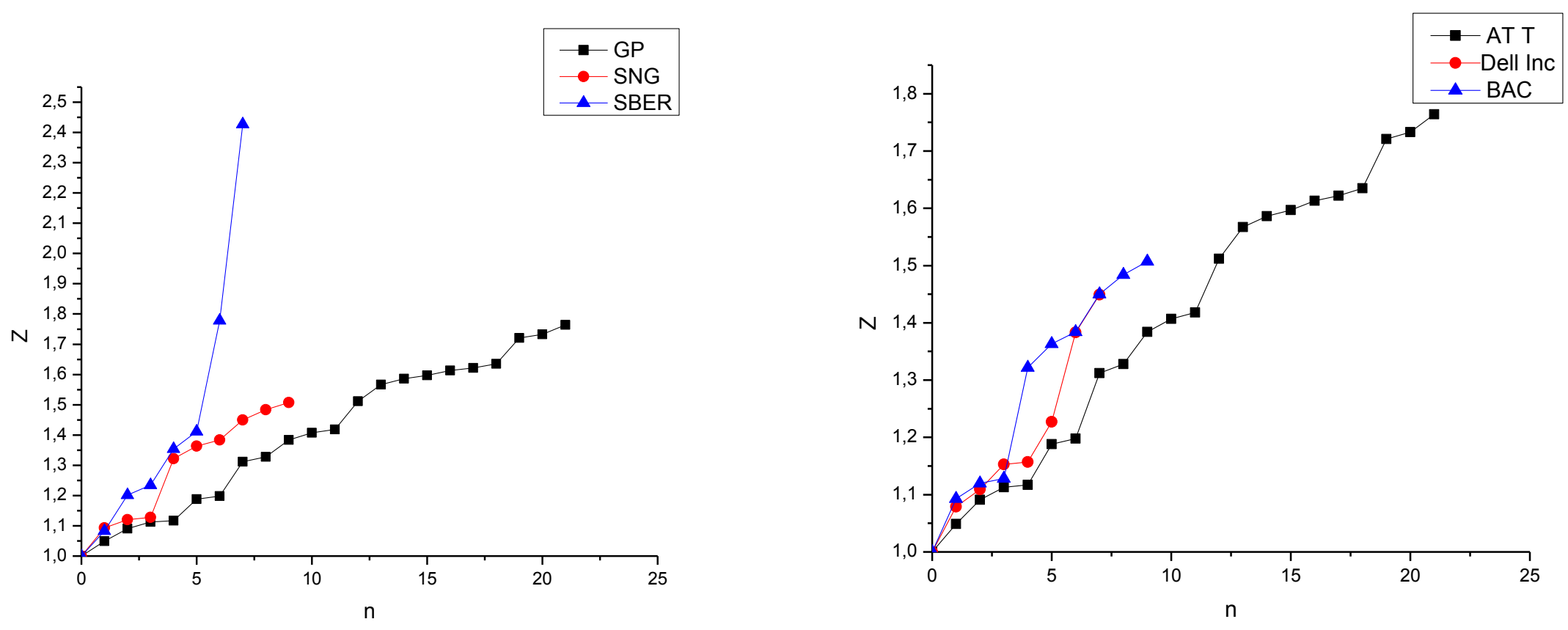

Figure 2 Profitability when using NSW algorithm within the period since January 1, 2009 till December 31, 2009 for stocks of Sberbank (SBER), Gazprom (GP) and SurgutNefteGaz (SNG) (a) and for AT&T, Dell Inc. and the Bank of America (b); and risk levels are $\alpha_1 = \alpha_2 = 0.05$.

Average fraction of decisions made (number of purchase and sale acts towards length of purchase and sale field) were in the interval 0.01...0.025. Daubechies wavelets were optimal for the stocks of Russian emitters and stocks of USA companies.

### 3. Optimization of security portfolio on the basis of NSW algorithm

Let $z_i(t)$ be the cost of instrument $i = 1..M$ at the point of time $t$ in accordance with abovementioned decision algorithm. Let $x_i(t) = \ln z_i(t+\tau_0)/z_i(t)$ be logarithmic earning power and let's define average value $X_i$ and two-point correlation function at the point of time $t > T_1$:

$$X_i = \frac{1}{T_1}\int_{t-T_1}^{t} x_i(t)dt$$

$$\lambda_{ij} = \frac{1}{T_1}\int_{t-T_1}^{t} (x_i(t) - X_i)(x_j(t) - X_j)dt$$

Lower estimate of interval $T_1$ is maximum value per $i$ of distribution quasi-stationarity with predetermined risk level $\alpha_1$ (fig. 1). Nondiagonal elements $\lambda_{ij}$ take account of intercorrelation of profitability of instruments $i, j$ in accordance with NSW algorithm.

If $n_i(t)$ is fraction of instrument $i$ in parcel at point of time $t$, then in Gaussian approximation average profitability of the parcel is $Z(t) = \sum_{i=1}^{M} n_i(t)X_i(t)$ and its expected mean square is $y^2(t) = \sum_{j=1}^{M}\sum_{i=1}^{M} n_i(t)n_j(t)X_i(t)X_j(t)$.

Moments of higher order $\lambda_{i...j}$ can be set aside, as far as by estimate for all instruments under consideration the following is right: $\left|\frac{\lambda_{i...j}}{\sqrt{\lambda_{ii}^2 .. \lambda_{jj}^2}}\right| \approx 0.002...0.004 << 1$.

Let's define the most effective parcel at point of time $t$, as a set of instruments $i$ with fraction $n_i(t)$, which maximizes value of probability

$P(\theta) = \int_{\theta Z(e)}^{\infty} \psi(z, y, s)ds \rightarrow extr$ under limitations $\sum_{i=1}^{M} n_i \leq 1$.

The parameter $\theta > 0$ implements the compromise between maximum earning power and minimum risk. When $\theta \rightarrow 0$ maximum of earning power is preferable, and when $\theta \rightarrow 1$ risk minimum is preferable. Thus, the task about the most effective portfolio is the typical task of nonlinear optimization. Necessary and sufficient condition of extremum existence is provided with Kun-Takker theorem. To solve the problem we shall use gradient procedure. It is possible to take $n_i^0 = 1/M$ as initial points.

In Figure 3 there is relation between the average profit and the share fraction in parcel against the time for optimal portfolio of AT&T Inc, Dell Inc. and the Bank of America over a period January 1, 2009 to June 30, 2009. We use Daubechies wavelets. As opposed to Figure 2, cost of the parcel is represented at point of time, when decision about redistribution of instruments in the parcel is made.

Trading was carried out in computer-aided mode without readjustment of the algorithm over a period of a half year. Only risk levels $\alpha_1, \alpha_2$ and compromise parameter $\theta$ between profitability and minimum risk were pre-set as the input.

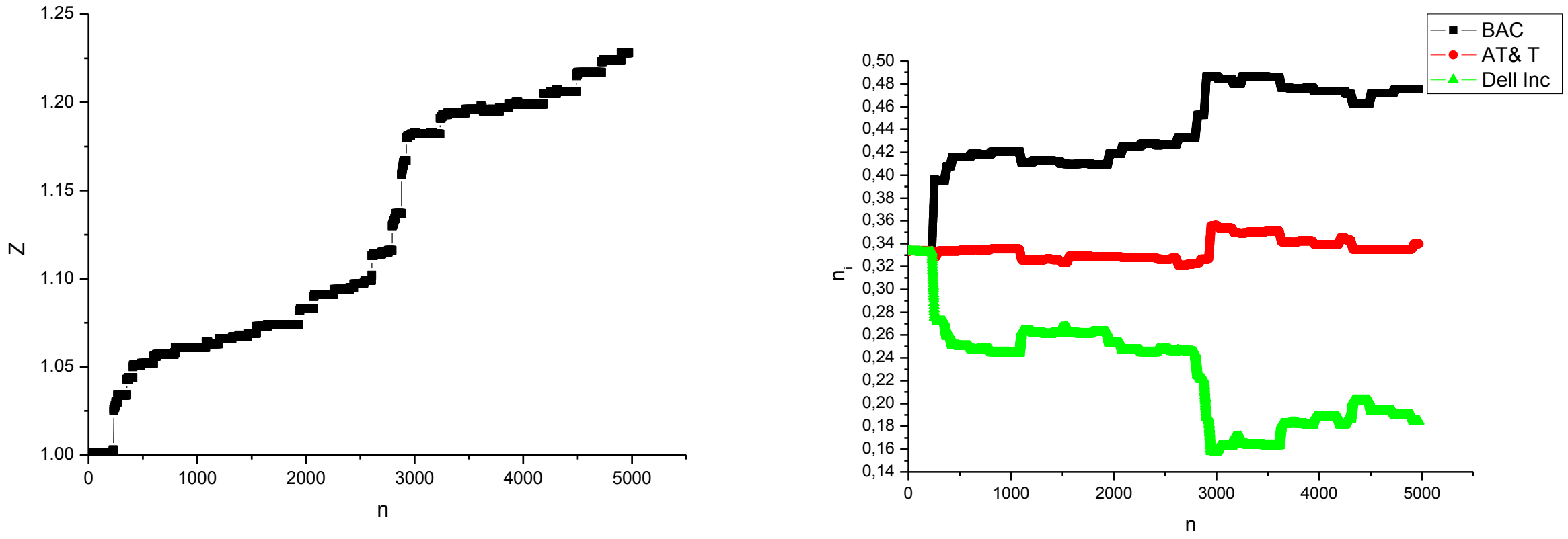

Figure 3. a) relation between average profitability of parcel of AT&T Inc, Dell Inc. and the Bank of America and time. Risk levels $\alpha_1 = \alpha_2 = 0.05$, and compromise parameter $\theta = 1/4$; b) share fraction in optimal parcel depending on time.

In ordinary moment the parcel is shared equally between AT&T Inc, Dell Inc and the Bank of America. However, in the process of trading according to NSW algorithm, redistribution of the parcel in favor of the Bank of America for the account of Dell Inc. takes place. Accumulated profit of the parcel is less than in case of dealing in the Bank of America shares (Figure 2), but it is achieved with the less risk.

## 4. Discussion about the results

In table 1 the NSW algorithm is compared with the following well-known algorithms: Price Channel, Bollinger Bands (BB), Moving Average Convergence/Divergence (MACD) and Relative Strength Index over a period of January 1, 2009 to December 31, 2009.

Parameters of standard algorithm were chosen to the effect that maximum earning power is provided. In this context, when comparing the algorithms, the well-known models had the advantage a priori; i.e. the parameters were chosen according to the process, which had been already realized (in reality it is natural that "future" is unknown).

Proposed NSW algorithm was implemented as close as possible to reality; predictions and control were fulfilled on the basic of knowing the previous system status only. Advantages of proposed algorithm are essentially for all financial instruments.

Table 1 Comparison of profitability of various algorithms, January 1, 2009 to December 31, 2009

| | PC | BB | MACD | RSI | NSW |
|---|---|---|---|---|---|
| The Bank of America | 1.596 | 1.631 | 1.273 | 1.683 | 1.852 |
| Dell Inc | 1.131 | 1.377 | 1.311 | 1.185 | 1.524 |
| AT&T Inc | 1.433 | 1.431 | 1.254 | 1.439 | 1.721 |

In table 2, comparison of algorithms was carried out in poor for the Bank of America market conditions. There was significant negative trend with strong volatility. Standard algorithms turn out to be unprofitable. NSW algorithm allows receiving of small, but stable profitability.

Table 2. Comparison of profitability of various algorithms for significant negative trends with great volatility. The Bank of America, May 1, 2010 to June 30, 2010.

| | PC | BB | CC | OC | NSW |
|---|---|---|---|---|---|
| The Bank of America | 0.99 | 0.908 | 0.918 | 0.872 | 1.078 |

The results analogous to those represented in tables 1 and 2 were received in automatic trading with other financial instruments, particularly with futures of Brent (BR-6.10, BR-7.10, BR-8.10, BR-9.10).

Thus, non-Markov generalization of Ito equation in wavelet-space enables implementation of “short” predictions of the most probable path of price change and statistics in the neighborhood of this path. Jointly with stationary distribution it enables to optimize the system of automatic trading.

## 5. Reference literature